\documentclass[pdflatex,sn-mathphys-num]{sn-jnl}

\usepackage{graphicx}%
\usepackage{multirow}%
\usepackage{amsmath,amssymb,amsfonts}%
\usepackage{amsthm}%
\usepackage{mathrsfs}%
\usepackage[title]{appendix}%
\usepackage{xcolor}%
\usepackage{textcomp}%
\usepackage{manyfoot}%
\usepackage{booktabs}%
\usepackage{algorithm}%
\usepackage{algorithmicx}%
\usepackage{algpseudocode}%
\usepackage{listings}%

\theoremstyle{thmstyleone}%
\theoremstyle{thmstyletwo}%

\theoremstyle{thmstylethree}%

\begin{document}

\title[THE GALACTIC HALO ROTATION BY WEYL INCORPORATED GRAVITY]{THE GALACTIC HALO ROTATION BY WEYL INCORPORATED GRAVITY}


\author[1]{\fnm{Asghar} \sur{Qadir}}\email{asgharqadir46@gmail.com}
\equalcont{These authors contributed equally to this work.}

\author[2]{\fnm{Ashmal} \sur{Shahid}}\email{shahidashmal@gmail.com}
\equalcont{These authors contributed equally to this work.}

\author*[3]{\fnm{Noraiz} \sur{Tahir}}\email{noraiz.tahir@sns.nust.edu.pk}

\affil[1]{\orgdiv{Pakistan Academy of Sciences}, \orgaddress{\street{Sector G-5/2}, \city{Islamabad}, \postcode{44000}, \state{Islamabad}, \country{Pakistan}}}

\affil[2]{\orgdiv{Department of Mathematics}, \orgname{School of Natural Sciences (SNS), National University of Sciences and Technology (NUST)}, \orgaddress{\street{Sector H 12}, \city{Islamabad}, \postcode{44000}, \state{Islamabad}, \country{Pakistan}}}

\affil*[3]{\orgdiv{Department of Physics \& Astronomy}, \orgname{School of Natural Sciences (SNS), National University of Sciences and Technology (NUST)}, \orgaddress{\street{Sector H 12}, \city{Islamabad}, \postcode{44000}, \state{Islamabad}, \country{Pakistan}}}


\abstract{A modification of the Einstein-Hilbert Lagrangian by introducing a coupling between the Weyl tensor and the stress-energy tensor was proposed to explain flat galactic rotation curves without the exotic (non-baryonic) dark matter (DM) \cite{qadir2019}. The proposed coupling constant was previously determined by fitting the rotational velocities of the Milky Way and M31 modeled with constant density, yielding the same coupling constant for both \cite{bilal2023, usama2025}. In this work, we have modified the formalism for a variable density by modeling the galactic systems with realistic, spherically symmetric and radially varying density profiles for the baryonic matter and this analysis is applied to seven edge-on spiral galaxies of the local cluster \cite{M31, M33, M81, M82, NGC5128, NGC4594, M90} and the Milky Way.}




\maketitle

\section{Introduction \label{intro}}
One of the most striking observations in galactic dynamics is the discrepancy
between the predicted and observed rotational velocities of galaxies.
According to the standard theories of gravity, the rotational velocity of the
galaxies should decrease sharply at large radii where visible matter becomes
sparse. However, observations of their rotation curves remain nearly flat out
to very large distances \cite{Rubin1980, Rubin1983, rubin2000}. Other dynamic
considerations had already led Zwicky \cite{zwicky1933} to propose the
existence of DM, but this evidence was much stronger. Rubin's investigation
was extended to galactic clusters \cite{tremaine1979,peebles1984} providing
yet stronger evidence. The observations of the cosmic microwave background
(CMB) had already provided minimum and maximum values for baryonic matter in
the Universe according to the standard model of particle physics (SMpp). The
observations required a value well beyond the limit of the baryonic matter
\cite{christian2018}. This has led to various suggestions for exotic
(non-baryonic) DM, but there is no direct evidence for any of the proposed
candidates. Nevertheless, CMB observations also indicate that $\simeq
5\%$ of the Universe should be made up of baryons (the usual protons and
neutrons), but observations of the luminous parts of the galaxies show
only half of these baryons, this is the ``missing baryon problem''. Since the baryons are dark, we call this the baryonic DM. It is proposed that a significant fraction of this baryonic DM is
present in the galactic halos \cite{planck2016baryons, planck2020, gunn1996,
gerhard1996, frasermcKelvie2011, nicastro2018}.

For the non-baryonic DM an alternative suggestion was that the standard
law of gravity should be modified instead of looking for other forms of
matter. The first such suggestion came from Milgrom \cite{milgrom1983}, who
proposed the modification of {\it Newton's} law by inserting a Yukawa-like
term to damp gravity at large distances, Modified Newtonian Dynamics (MOND).
It was not able to explain the dynamics at different scales, especially of
single galaxies and clusters, or provide for the formation of structure in
the early Universe \cite{Famaey2012, Sanders1999, Angus2008, McGaugh2015,
Skordis2020, Bekenstein2004, Llinares2008}. Most of all, the damping term was
totally ad-hoc and was embedded in an obsolete Newtonian framework, which
could not be converted to General Relativity (GR) \cite{Skordis2020}. Apart
from galactic dynamics, arguably the most outstanding problem of fundamental
physics is the incompatibility of GR and Quantum Theory. In particular, the
Renormalization Group Equation of 't Hooft and Veltman demonstrated that
Dirac quantization of GR produced a {\it non-renormalizable} theory
\cite{roberto2023}, leading to the well-known ``Quantum Gravity (QG)''
problem. To avoid these separate problems various ad-hoc modifications of GR
involving arbitrarily many new parameters have been proposed
\cite{milgrom1983, sotiriou2010, cai2016, capozziello2011, harko2011}.

Qadir and Lee took the view that there must be a sound physical basis for any
modification of the highly successful GR, and it must be minimal, i.e., it
should remain geometric and involve only one free parameter to explain the
discrepancies of galactic dynamics at all scales. Further, it should also
provide a base for solving the QG problem. In 2019, they proposed an explicit
interaction term between matter and the gravitational field $\lambda
\mathbf{T.C.T}$, where $\lambda$ is a new coupling constant, $\mathbf{T}$ is
the stress-energy tensor, and $\mathbf{C}$ is the Weyl tensor, which
represents a pure gravitational field \cite{qadir2019}. This idea was
inspired by the Feynman vertex representing a similar explicit interaction
between the source (an electron) and the electrodynamic field in Quantum
Electrodynamics (QED), $A_{\mu}j^{\mu}$. In QED the electron is given by a
spin-half spinor, which comes twice over in the current $j^{\mu}$ and the
field, $A_{\mu}$, which appears singly, while in QG the source would be
represented by the rank-two tensor {\bf T}, which comes twice, and the
gravitational field by the rank-four tensor {\bf C}, which comes once. As
QED with this source term is renormalizable, it can be hoped that so would
this modified QG. It was called {\it Modified Relativistic Dynamics} (MORD).

Previously, MORD was tested by checking whether a single value of the
new coupling constant $\lambda$ could reproduce the rotational velocity at
the outer rim of two galaxies, the Milky Way and M31, incorporating only the
baryonic DM and not any postulated non-baryonic DM, by assuming a
simple-minded model in which both the galaxies were represented as a constant
density sphere with a peak density of the baryonic matter from the core to
the edge of the galaxy \cite{bilal2023, usama2025}. In that study, a single
value of $\lambda$ was indeed found to fit the rotational velocity values for
both galaxies. This approach was inherently limited, i.e., it neglected the
radial variation of the galactic halo density, and treated the galaxies as
idealized, uniform objects. The aim of this paper is to completely modify the previous formalism of a constant density case to a variable density case, where $\rho' \neq 0$, and take the next step forward by generalizing the baryonic DM component to spherically symmetric, radially varying density profiles for the galactic halos of \textit{eight} spiral galaxies \cite{sofue2020rotation, russeil2017, M31, M33, M81, M82,
NGC5128, NGC4594, M90}.

This extension is conceptually significant because it will allow us later to test whether the universality of $\lambda$ persists under physically motivated halo structures at different radii, rather than only at the rim. By moving from a toy model to a realistic halo description, we not only refine the numerical estimate of $\lambda$ but also provide a
more robust and physically meaningful assessment of MORD across multiple
spiral galaxies. We stress that while more realistic baryonic distributions,
such as double exponential stellar disks combined with bulge components, are
commonly used to model luminous matter, these structures are intrinsically
non-spherical, and would require extension of the formalism to two or more
variables.

The plan of the paper is as follows: in Section \ref{mordandmw}, we will
briefly explain the Weyl modified Einstein field equations for varying
spherically symmetric density profiles and demonstrate how the value of the
coupling constant $\lambda$ is obtained for the Milky Way galactic halo. In
Section \ref{otherspirals}, we will use the analysis for seven other spiral
galaxies. Finally, in Section \ref{results}, the obtained results will be
discussed.
\section{Review of MORD and its Application to the Milky Way Halo \label{mordandmw}}

\subsection{Brief Review of MORD \label{brief review}}
The Weyl-modified Einstein-Hilbert Lagrangian is \cite{qadir2019}
\begin{equation}
   \mathcal{L} = \sqrt{-g} \left( R - 2\Lambda - \kappa T + \lambda C_{\alpha\mu\beta\nu} T^{\alpha\beta} T^{\mu\nu} \right),
   \label{modifiedlagrangian}
\end{equation}
where $\sqrt{-g}$ is the determinant of the metric, $\kappa = 8\pi G/c^4$ is the coupling constant for matter, where $G$ is the Newton's gravitational constant, $c$ is the speed of light. This leads to the Weyl incorporated Einstein field equation (WIFE)
\begin{equation}
R_{\mu \nu} - \frac{1}{2} g_{\mu \nu} R + g_{\mu \nu} \Lambda = \kappa T_{\mu \nu} + \lambda I_{\mu \nu}\, ,
\end{equation}
where $I_{\mu \nu}$ is the interaction term given by
\begin{align}
I_{\mu\nu} = &\frac{1}{4} \left( -g_{\alpha\beta}g_{\rho\mu}g_{\sigma\nu}- g_{\rho\sigma}g_{\alpha\mu}g_{\beta\nu} - g_{\alpha\sigma}g_{\rho\mu}g_{\beta\nu} - g_{\rho\beta}g_{\alpha\mu}g_{\sigma\nu} \right) \square \left( T^{\alpha\beta} T^{\rho\sigma} \right)\nonumber\\ &+\frac{1}{6} \left( g_{\alpha\beta}g_{\rho\sigma} - g_{\rho\beta}g_{\alpha\sigma} \right) \left( g_{\mu\nu} \square - \nabla_\mu \nabla_\nu \right) 
\left( T^{\alpha\beta} T^{\rho\sigma} \right).
\end{align}
For a spherically symmetric metric, $ds^2 = e^{\mu(r)} dt^2 - e^{-\nu{(r)}} dr^2 - r^2 d\theta^2 - r^2 \sin^2\theta d\phi^2 $ the modified Einstein tensor is given by
\begin{align}
G_{tt} & = e^{\nu(r) - \mu(r)} \left( \frac{r \mu'(r) + e^{\mu(r)} - 1}{r^2} \right) \nonumber \\ & =  \kappa e^{\nu(r)} \rho(r) + \lambda e^{\nu(r) - \mu(r)} \left( \frac{1}{2r} \right. \left( \frac{r \nu'(r)^2}{\rho(r)^2} - 2r \nu'(r) \frac{\rho'(r)}{\rho(r)} \right. \nonumber \\ & \left. - 4r \frac{\rho'(r)^2}{\rho(r)^2} \left. - 4r \frac{\rho(r) \rho''(r)}{\rho(r)^2} + 2r \mu'(r) \frac{\rho''(r)}{\rho(r)} - 8 \frac{\rho(r) \rho''(r)}{\rho(r)^2} \right) \right),  
\label{gtt}
\end{align}
and 
\begin{equation}
G_{rr} = -\frac{1}{r^2} \left( -r \nu'(r) + e^{\mu(r)} - 1 \right) = -\kappa e^{\mu(r)} P + \lambda \frac{\nu'(r)^2}{2 \rho(r)^2}.
\label{grr}
\end{equation}
Here, $\rho(r)$ is the density distribution of the galaxy. Here 
\begin{equation}
\mu(r) = F(r) - \ln \left[ \left\{ \frac{r}{1 - \lambda r c^4 \rho(r) \rho'(r)} \left( \int \frac{e^{F(r)}}{r^2} dr - \kappa c^2 \int \rho(r) e^{F(r)} dr \right) \right\} \right]
.
\label{mu}
\end{equation}
Also,
\begin{equation}
F(r) = \frac{1}{2} \int \frac{4 + \lambda r (r \rho^2 \nu'(r)^2 - 2 r \rho(r) \rho'(r) \nu'(r) - 2 r \rho(r)^2 - 2 r \rho(r) \rho'(r) - 8 \rho(r) \rho'(r))}{r (1 - \lambda r \rho(r) \rho'(r))} dr,
\label{f}
\end{equation}
and
\begin{equation}
\nu'(r) = \frac{1 - \sqrt{1 - 2 \lambda \rho(r)^2 c^4 \left( e^{\mu(r)} - 1 \right)}}{r \lambda \rho(r)^2 c^2}.
\label{nu}
\end{equation}
Similarly, the rotational velocity is given by
\begin{equation}
v(r)^2 = \frac{r\nu'(r)}{2} =\frac{1 - \sqrt{1 - 2 \lambda \rho(r)^2 c^4 \left( e^{\mu(r)} - 1 \right)}}{2r \lambda \rho(r)^2 c^2}. 
\label{v}
\end{equation}

Notice that eqs. (\ref{f}) - (\ref{v}) are implicit equations. In order to make them explicit, we first take the anti-log of eq. (\ref{mu}), and then we take the derivative of the resulting equation with respect to $r$. The
differential equation obtained is
\begin{align}
\frac{d\mu(r)}{dr} = & \frac{dF(r)}{dr} - \left( \frac{r}{1 - \lambda c^4 r \rho(r) \rho'(r)} \right) 
\left[\left( \frac{\lambda r c^4 (\rho(r) \rho''(r)+\rho'(r)^2)+1}{r^2} \right) \right. \nonumber \\ &\left. +e^\mu(r)\left(\frac{1-\kappa r^2c^2\rho(r)}{r^2}\right)\right], 
\label{dmu/dr}
\end{align}
where 
\begin{align}
    \frac{dF(r)}{dr} = \frac{1}{2} & \left( \frac{4 + \lambda r c^4 \left( r \rho(r)^2 \nu'(r)^2 - 2 r \rho(r) \rho'(r) \nu'(r) \right. }{r} ... \right. \nonumber \\ & \left. ... \frac{\left.- 2 r \rho'(r)^2 - 2 r \rho(r) \rho''(r) - 8 \rho(r) \rho'(r) \right)}{\left( 1 - \lambda r c^4 \rho(r) \rho'(r) \right)} \right) \, .
\label{df/dr}
\end{align}
Substituting eqs. (\ref{nu}) and (\ref{df/dr}) in eq. (\ref{dmu/dr}), we obtain the differential equation for $\mu(r)$
\begin{align}
    \frac{d\mu(r)}{dr} = & \frac{1}{2r \left(1 - \lambda r c^4 \rho(r) \rho'(r)\right)} \left[ 4 - 2 e^{\mu(r)} \left(1 - \kappa r^2 c^2 \rho(r)\right) + \right. \nonumber\\ & \left. \frac{\left(1 - \sqrt{1 - 2 \lambda \rho(r)^2 c^4 \left(e^{\mu(r)} - 1\right)}\right)^2}{4 \lambda \rho(r)^2} - 2 r^2 \left(\lambda r c^4 \left(\rho(r) \rho''(r) + \rho'(r)^2 \right) + 1 \right) \right. \nonumber\\ & \left. - \frac{r \rho'(r) c^2 \left(1 - \sqrt{1 - 2 \lambda \rho(r)^2 c^4 \left(e^{\mu(r)} - 1\right)}\right)}{\rho(r)} - \lambda r c^4 \left(2 r \rho'(r)^2 \right. \right. \nonumber \\ & \left. \left. + 2 r \rho (r) \rho''(r) + 8 \rho(r) \rho'(r) \right) \right].
    \label{DE}
\end{align}

\subsection{The Milky Way Halo Rotational Velocity \label{mwrotation}}
The rotational curve of the Milky Way plays a central role in our understanding of galactic dynamics and the mass distribution of spiral galaxies. Unlike external galaxies, for which rotation curves can be directly measured through Doppler shifts across the disk, the Milky Way rotation curve must be reconstructed using indirect kinematic tracers. The Galactic rotation curve remains approximately flat from the Solar radius $R\sim 8$ kpc out to large galactocentric distances $R \gtrsim 50-100$ kpc \cite{sofue2020rotation, russeil2017, yang2016}. In the inner Galaxy, the rotation curve is constrained by stellar kinematics, molecular and atomic gas observations, and maser measurements, while at larger radii it is inferred from halo stars, globular clusters, satellite galaxies, and stellar streams \cite{sofue2020rotation, russeil2017, yang2016}. These independent tracers consistently indicate that the circular velocity does not decline in a Keplerian manner, as expected from the observed distribution of luminous matter alone. Instead, the nearly constant rotational velocity implies that the enclosed mass continues to grow with radius, suggesting a substantial halo component that dominates the Galactic potential at large radii.

The physical nature of this halo component is expected to consist of virial clouds, which are observed in the halos of nearby edge-on spiral galaxies using the \textit{Planck} CMB data \cite{depaolis1994, depaolis1995, tahir2023galactic, qadir2019virial, tahir2019, tahir2019seeing}. Theoretical and observational studies suggest that such virialized baryonic structures can carry a significant mass fraction $\approx 90 \%$  of the DM halo while remaining consistent with the total cosmic baryon fraction and observational constraints on halo baryons \cite{tahir2023planck, tahir2022rksz, tahir2025ame, depaolis2025m90, tahir2024hvcs}. Three widely adopted spherically symmetric density profiles are used to explain the distribution of the virial clouds in the halos \cite{tahir2023galactic}, and we assume that this is also the case for the Milky Way halo. 

The first two density profiles appear to be singular at the galactic centre. This is not a matter of concern as they apply outside the core radius and do not start at the centre of the Galaxy. When using any one of them, it is common to choose some core radius that is seen as convenient. Since we are dealing with all three at the same time, we spell out our procedure and provide our justification for doing so. We take the core of the visible part of the galaxy, inside which visible baryonic matter dominates over baryonic DM. The matter inside the core is taken to be provided, not by the luminosity but by the rotational velocity at the edge of the core, $r_c$. As such the baryonic DM is included in the core density $\rho_c$. These values are independent of the specific model but would vary from galaxy to galaxy.

\begin{figure}
    \centering
    \includegraphics[width=0.9\linewidth]{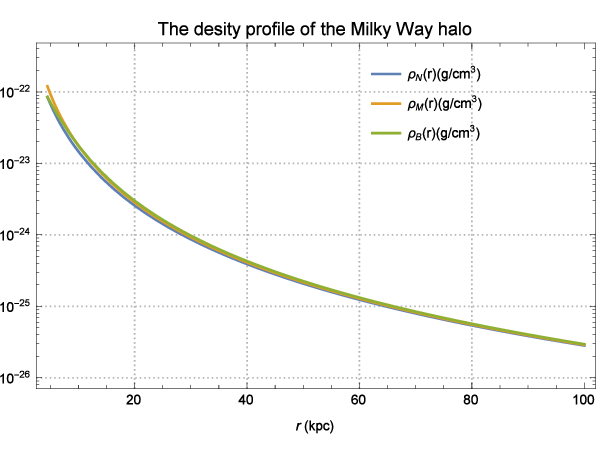}
    \caption{The comparison of the three density profiles: the NFW (Blue Curve); Moore (Orange curve); and Burkert (Green Curve) profiles of the Milky Way halo up to $100$ kpc \cite{tahir2023galactic}, for $r_c= 4.5$ kpc, and $\rho_c=6.99 \times 10^{-23}~{\rm g/cm^3}$ \cite{shin2007}.
\label{milkywaydensityprofiles}}
\end{figure}

\begin{figure}
    \centering
    \includegraphics[width=0.9\linewidth]{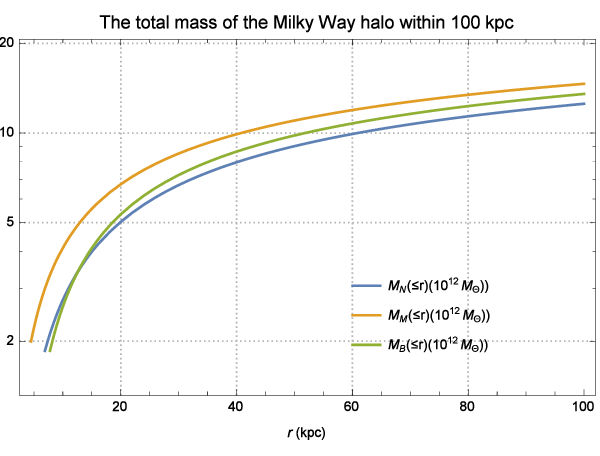}
  \caption{The total mass of the Milky Way halo within $100$ kpc obtained by using eq. (\ref{massequation}). The blue curve represents the mass profile from the NFW model, the orange curve represents the mass profile from the Moore model, and the green curve represents the mass profile obtained from the Burkert model, respectively.
\label{massofmilkyway}}
\end{figure}

For the Milky Way galaxy $r_c= 4.5$ kpc, and $\rho_c=6.99 \times 10^{-23}~{\rm g/cm^3}$ \cite{shin2007}. In Fig. \ref{milkywaydensityprofiles} we give the three density profiles, the NFW (Blue Curve); Moore (Orange curve); and Burkert (Green Curve) for the Milky Way halo up to $100$ kpc. The total density within $100$ kpc in the case of the NFW profile is $3.83 \times 10^{-25}~{\rm g/cm^3}$, in the case of the Moore profile is $5.61 \times 10^{-25}~{\rm g/cm^3}$, and in the case of the Burkert profile is $4.42 \times 10^{-25}~{\rm g/cm^3}$, and the average density for all the three models within $100$ kpc came out to be $3.21 \times 10^{-27} ~ {\rm g/cm^3}$. One can clearly see that the obtained densities from the three profiles are comparable with each other.
The total mass of the Milky Way halo can be estimated by using the relation 
\begin{equation}
    M(\leq r) = \int_0^r 4\pi q^2 \rho^{N,M,B}(q) dq.
    \label{massequation}
\end{equation}
Here, $\rho^N$, $\rho^M$ and $\rho^B$ are the three density distributions \cite{tahir2023galactic}. The obtained mass profile is shown in Fig. \ref{massofmilkyway}. The total mass within $100$ kpc for the three models came out to be $1.86 \times 10^{12}~M_\odot$.
To fit the rotational velocity of the Milky Way, we consider the three density distributions one by one and numerically solve eq.~(\ref{DE}) for the metric function $\mu(r)$ for each density profile. The equation is treated as a second-order ordinary differential equation (ODE) and integrated radially outward from the galactic center by imposing the regularity condition $\mu(r=0)=\mu_0$, where $\mu_0$ is finite. The numerical integration of eq.~(\ref{DE}) is performed using an adaptive step-size Runge--Kutta (RK) algorithm, ensuring stability and accuracy over the full radial domain up to $r = 100~\mathrm{kpc}$.

For a guess value of $\lambda$, the numerical solution for $\mu(r)$ is substituted into eq.~(\ref{v}) to compute the corresponding rotational velocity profile. The value of $\lambda$ is then iteratively refined using a shooting method until the modeled rotational velocity matches the observed Milky Way halo velocity at the outermost radius. Convergence is achieved when successive iterations in $\lambda$ result in variations in the fitted velocity smaller than the prescribed numerical tolerance.

We find that a common value $\lambda = 6.9546 \times 10^{-18}~{\rm km^2\,s^4\,kg^{-2}}$, together with $\mu_0 = 1.0458 \times 10^{-31}$, provides the best fit to the Milky Way rotation velocity value at $100~\mathrm{kpc}$ for all three halo models and are given in Table \ref{parametersforeachgalaxy} (see column 5-7). Here, the quoted rotation velocity refers to the circular velocity inferred from the Milky Way rotation curve, which characterizes the gravitational potential at large radii, and should not be confused with the bulk rotational motion of the halo itself, which observationally is expected to be much smaller.

The stability of this result under variations of the initial conditions and halo profiles, as well as its extension to other spiral galaxies, is discussed in detail in Section~\ref{otherspirals}. Notably, the inferred value of $\lambda$ is significantly smaller than that obtained previously for the constant-density toy model in Ref.~\cite{usama2025}, highlighting the importance of adopting radially varying density distributions.

\section{Halo Rotational Velocity of other Edge-on Spiral Galaxies \label{otherspirals}}
We now extend our analysis to seven other galaxies in the local group, M31, M33, M81, M82, NGC5128, NGC5494, and M90, whose halo rotations have been studied using CMB Planck data and other observations \cite{tahir2023galactic, tamm2012, depaolis2025}. Along with the Milky Way, the rotational velocity curves of these nearby galaxies offer essential constraints on mass distribution and gravitational potential at galactic scales.

These galaxies span a range of morphological types and environmental conditions, yet each exhibits evidence for a substantial dark matter halo. M31, the nearest massive spiral, shows a flat rotation curve extending beyond 100 kpc in ${\rm H_{I}}$, stellar kinematics, satellite dynamics, and CMB data, indicating a dominant halo component. M33, a lower-mass spiral, has a slowly rising rotation curve that remains below the plateau typically seen in more massive galaxies—yet its velocities still exceed predictions from luminous matter alone, underscoring the importance of halos even in low-mass systems.

M81 displays a classical flat rotation curve extending well beyond its stellar disk, while its companion M82 shows more disturbed kinematics due to tidal interactions and outflows, though high velocities in its outer regions still point to a stabilizing halo. Beyond spirals, NGC5128—an early-type galaxy with a prominent dust lane—exhibits significant rotational support at large radii from planetary nebulae and globular cluster kinematics, implying a massive halo despite its elliptical morphology. Similarly, NGC5494, a disk galaxy with extended ${\rm H_{I}}$ emission, shows a nearly flat rotation curve past its optical radius, and M90, a Virgo Cluster spiral, retains rotation speeds higher than baryonic predictions alone can explain, even in the presence of environmental effects like ram-pressure stripping \cite{M31, M33, M81, M82, NGC5128, NGC4594, M90}.

For these galaxies, we again consider the three density distributions previously used for the Milky Way. The core densities (column 2), radii (column 3), and the total virial mass within $100$ kpc (column 4) for each galaxy are shown in Table \ref{parametersforeachgalaxy}. The obtained values of mass for the corresponding core density and radius are in agreement with the values presented in the literature \cite{hartwick1974mass, tahir2023galactic}. To fit the estimated value of ${\rm v}_{rot}(r)$ at $100$ kpc to the observed value for these galaxies, we follow the same procedure used for the Milky Way halo. The best-fit rotational velocity values for all other seven galaxies are obtained for $\lambda = (6.95456 \pm 0.00012) \times 10^{-18} ~{\rm km^2 s^4/kg^2}$, and $\mu_o = (1.0458 \pm 0.0005) \times 10^{-31}$. The fitted values of ${\rm v}_{rot}$ at $100$ kpc are also shown in Table - \ref{parametersforeachgalaxy} (see column 5-7).

{\bf \begin{table}[t]
\caption{Estimated parameters for the eight galactic halos.}\label{parametersforeachgalaxy}
\begin{tabular}{@{}lllllllll@{}}
\toprule
Galaxy & $\rho_c$  & $r_c$  & $M_{h}(\leq r)$ & ${\rm v}_{rot}^{N}$  & ${\rm v}_{rot}^{M}$  & ${\rm v}_{rot}^{B}$ & ${\rm v}_{rot}^{obs}$ \\
    &$({\rm g cm^{-3}})$ & (kpc) & $10^{12} M_{\odot}$ & km/s &  km/s &  km/s  & km/s \\
\midrule
        Milky Way & $6.99 \times 10^{-23}$ & $4.5$ & $1.86$  & $156.78 \pm 0.45$ & $152.96 \pm 0.32 $ & $159.34 \pm 0.42 $ & $ 150 \pm 10 $ \\ 
        
        M31 & $2.27 \times 10^{-26}$ & $3.2$ & $1.40 $ & $231.65 \pm 0.45 $ & $229.92 \pm 0.32 $ & $232.34 \pm 0.42 $ & $ 225 \pm 10 $ \\  
      
        M33 & $4.40 \times 10^{-20}$ & $0.50$ & $32.0$ & $121.91 \pm 0.64 $ &  $120.84 \pm 0.22 $ & $122.32 \pm 0.25 $ & $ 120 \pm 5$  \\
     
        M81 & $5.87 \times 10^{-21}$ &  $1.53$ & $1.30$ & $256.14 \pm 0.66 $ & $255.72 \pm 0.75 $ & $257.41 \pm 0.76 $   &  $250 \pm 15 $  \\
     
        M82 & $3.91 \times 10^{-24}$ & $7.5$ & $10.2$ & $251.52\pm 0.22 $ & $250.68 \pm 0.35 $ & $253.38 \pm 0.54 $  & $ 250 \pm 18 $ \\ 
      
        NGC5128 & $2.71 \times 10^{-22}$ & $6.4$ & $4.40$ & $151.91 \pm 0.23 $ & $149.72 \pm 0.25 $ & $153.41 \pm 0.24 $   & $ 150 \pm 20 $ \\
      
        NGC4594 & $1.12 \times 10^{-20}$ & $4.5$ & $630$ & $193.93 \pm 0.55 $ & $191.29 \pm 0.62 $ & $196.62 \pm 0.70 $ & $ 190 \pm 25 $  \\
     
        M90 & $2.09 \times 10^{-20}$ & $3$ & $350$ &  $ 225.54 \pm 0.62 $  &  $223.31 \pm 0.58 $  &  $ 227.56 \pm 0.61 $ & $ 220 \pm 15 $  \\
       
\botrule
\end{tabular} 
\textbf{Note:} We give the considered galaxy (column 1), the core density (column 2), the core radius (columns 3), and the total halo mass (column 4), respectively. We also give the estimated rotational velocity at $R_{\rm halo} = 100$ kpc for the adopted DM model in columns 5-7. We also give the observed rotational velocity up to $100$ kpc from the galactic center (see column 8) \cite{M31, M33, M81, M82, NGC5128, NGC4594, M90}.
\end{table}
}
\section{Results and Discussion \label{results}}
The observed flatness in the rotational velocity curves of the galaxies and clusters of galaxies has raised questions on the kinematics and dynamics of the galaxies. To explain the flatness in the rotational curve: $(i)$ the presence of the exotic DM matter; and $(ii)$ modification in the standard theory of gravity have been proposed. Despite great advances in the field of particle physics, there is no evidence of any DM candidate fitting in the SMpp. Hence one could try to solve the flatness in the rotational velocity curves of the galaxies by making minimal modification in the standard GR. 

In contrast to the modification in conventional particle DM models typically require extensive multi-parameter fitting to explain cosmological and astrophysical observations. One of the notable example is the series of papers which explore supersymmetric hybrid inflation models with gravitino DM (c.f. \cite{lazarides2021shifted, afzal2022hybrid, rehman2018gravity}). This multi-parameter fitting, involving scans over high-dimensional spaces to satisfy multiple observational constraints, which only provides ``consistency'' with the observed data and no actual prediction or testing, motivates the search for alternative explanations that achieve comparable or better agreement with data using fewer free parameters. 

Modified gravity approaches offer precisely such a parsimonious alternative. The very first modified gravity approach to explain the flat rotational curves of galaxies without invoking DM is MOND \cite{milgrom1983}. However, MOND's phenomenological success comes with challenges in covariant formulation and cluster-scale dynamics, motivating alternative modifications rooted in GR. Furthermore, a major challenge to MOND has emerged from the analysis of wide binary stars in Gaia DR3. A comprehensive study by Ref. \cite{banik2024wide} found that the relative velocities of widely separated binaries ($2-30$ kAU) are inconsistent with the MOND prediction, which expects a $\approx 20\%$ enhancement over Newtonian gravity due to the external field effect. Their analysis, which rigorously modeled the Galactic external field and population uncertainties, excluded MOND at a statistical significance of $16\sigma$ in favour of Newtonian dynamics \cite{banik2024wide, pittordis2025wide}. These challenges motivate the exploration of alternative single-parameter modified gravity theories rooted in a covariant framework.

Qadir and Lee's MORD \cite{qadir2019} proposed a modification to the standard Lagrangian by incorporating an interaction coupling constant $\lambda$ with a term involving the Weyl tensor and the stress-energy tensor, expressed as $\lambda C_{\mu\nu\rho\pi}T^{\mu\nu}T^{\rho\pi}$ whose purpose was to see whether a single unique value of $\lambda$ can account for the rotational velocity curves of galaxies by replacing the exotic DM by what we now feel should be called Weyl Incorporated Gravity (WIG), solving the outstanding DM problem with a \textit{single} new parameter. This was tested using a simplistic constant density model which \textit{did} have just the one value of the coupling for all galaxies considered \cite{bilal2023, usama2025}.

In the present work, we have modeled the galactic halos of eight spiral galaxies, modifying the previous framework for a variable density case to estimate the value of the coupling constant $\lambda$. For this purpose we adopt three widely used density profiles, the Navarrow-Frenk-White (NFW), Moore, and Burkert models, normally used for {\it all} DM in the halos, but here used only for the baryonic DM to test the robustness of the proposal by verifying that the choice of model
makes no difference to the results \cite{tahir2019, tahir2019seeing,
tahir2023galactic, depaolis1994,  depaolis1995, qadir2019virial, tahir2023planck, tahir2022rksz, tahir2025ame, depaolis2025m90, tahir2024hvcs}. We find that {\bf a tiny range}, $\lambda = (6.9546 \pm 0.00012) \times 10^{-18}~{\rm km^2\,s^4\,kg^2}$, consistently reproduces the observed halo rotational velocities at $r = 100$~kpc. For the Milky Way, the fitted rotational velocities span ${\rm v}_{\rm rot} \simeq 153$--$159$~km~s$^{-1}$, compared to the observed value $150 \pm 10$~km~s$^{-1}$, with an enclosed halo mass $M_h(\leq 100~{\rm kpc}) \simeq 1.0 \times 10^{12}~M_\odot$. For M31, we obtain $v_{\rm rot} \simeq 230$--$232$~km~s$^{-1}$ versus the observed $225 \pm 10$~km~s$^{-1}$, corresponding to a halo mass $M_h \simeq 1.4 \times 10^{12}~M_\odot$. In the case of M33, the modeled velocities ${\rm v}_{\rm rot} \simeq 121$--$122$~km~s$^{-1}$ agree with the observed $120 \pm 5$~km~s$^{-1}$, yielding $M_h \simeq 3.2 \times 10^{11}~M_\odot$. For M81 and M82, the fitted velocities lie in the ranges $256$--$257$~km~s$^{-1}$ and $251$--$253$~km~s$^{-1}$, respectively, consistent with the observed values $250 \pm 15$~km~s$^{-1}$ and $250 \pm 18$~km~s$^{-1}$, with inferred halo masses $M_h \simeq 1.3 \times 10^{12}~M_\odot$ and $1.0 \times 10^{11}~M_\odot$. Similarly, for NGC~5128, NGC~4594, and M90, the modeled rotational velocities at $100$~kpc fall within the observed ranges reported in the literature, with corresponding halo masses $M_h \simeq 4.4 \times 10^{12}~M_\odot$, $6.3 \times 10^{13}~M_\odot$, and $3.5 \times 10^{13}~M_\odot$, respectively. It is clearly seen that these mass and velocity estimates are consistent with the observed values (see Refs. \cite{tahir2019, tahir2023galactic}, and Table ~\ref{parametersforeachgalaxy}).

Before closing the paper, note that we have used spherical symmetry to explain the dynamics of the galactic halo to estimate the value of $\lambda$. However, to make a more realistic model of the galaxy, we should take into account the vertical component of the velocity, which is missing in the present geometry. The hope is that one can fit the complete rotational velocity curve and get a more robust model. Indeed, the chosen metric would change, and we may need to consider the Kerr geometry, or a slow rotation approximation of it \cite{rezzolla2001}, to incorporate the vertical component, as it accounts for the angular momentum effects \cite{misner1973, mcvittie1933}. This will be addressed separately later. 
 
As previously discussed, the problem of DM and QG may share a common origin \cite{qadir2019}. Addressing observational issues related to DM could provide insights into resolving fundamental difficulties in QG. Instead of assuming an indirect interaction, we have introduced a direct nonlinear coupling between matter and gravity, analogous to the interaction between electromagnetic sources and the electromagnetic field. The modified Lagrangian, given in eq. (\ref{modifiedlagrangian}), represents the minimal extension of the Einstein-Hilbert Lagrangian and is proposed as a potential solution to both problems with a single additional parameter. Naturally, the feasibility of this approach must first be tested against the DM problem before seeing if our WIG fits on the messier head of QG.

\section*{Acknowledgments}
NT would like to acknowledge support from the NUST flagship project NUST-24-41-78. AQ, AS, and NT would also like to acknowledge Prof. Viqar Husain for his suggestion on the work. We would also like to acknowledge the anonymous referees for the constructive comments.
\bmhead{Data Availability} No Data associated with the manuscript.

\bmhead{Declaration of Conflicting Interests} The authors declare that there are no conflicts of interest.

\end{document}